\documentclass[]{gPAA2e}

\usepackage{etex}
\usepackage{placeins}
\usepackage{textcomp}
\usepackage{booktabs}
\usepackage{multirow}

\usepackage[Bjarne]{fncychap}
\usepackage{color}
\usepackage{listings}
\usepackage{fp}
\usepackage{colortbl}

\usepackage{epigraph}

\usepackage{mathtools}
\usepackage{tikz}
\usepackage{tikz-uml}
\usetikzlibrary{arrows.meta}
\usetikzlibrary{positioning}
\usetikzlibrary{mindmap}
\usetikzlibrary{trees}
\usetikzlibrary{shapes}
\usetikzlibrary{shadows}
\usetikzlibrary{calc}
\usetikzlibrary{snakes}
\usetikzlibrary{decorations.markings}
\usetikzlibrary{fit}
\tikzumlset{fill class=gray!5}

\usepackage{pgfplots}
\usepackage{pgfplotstable}

\usepgfplotslibrary{statistics}

\usepackage{float}
\usepackage{newfloat}
\usepackage{caption}

\usepackage{subfig}
\usepackage{xspace}
\usepackage{ifthen}

\makeatletter
  \newcommand{\camelhyph}[1]{\@fterfirst\c@amelhyph#1\relax }
  \def\@fterfirst #1#2{#2#1}
  \def\c@amelhyph #1{%
   \ifthenelse{\equal{#1}\relax}{}{
     \ifnum`#1<91 \-#1\else#1\fi
    \expandafter\c@amelhyph
}}
\makeatother

\newcommand{\cclass}[1]{\texttt{\camelhyph{#1}}\xspace}

\newcommand{\percentright}[2][gray]{
  \begin{tikzpicture}
    \fill[#1] (0,0) rectangle (7 * #2,0.4);
    \FPeval{\result}{clip(#2*100)}
    \node[] at (7 * #2 + 0.35, 0.2) { \scriptsize \result  \%};
  \end{tikzpicture}
}

\newcommand{\percent}[2][gray]{
  \begin{tikzpicture}
    \FPeval{\result}{clip(#2*100)}
    \node[] at (-0.35, 0.2) { \scriptsize \result  \%};

    \fill[#1] (0,0) rectangle (7 * #2,0.4);
  \end{tikzpicture}
 }

\newcommand{\boxplot}[6][0.7]{%
\begin{tikzpicture}[rotate=180, yscale=#1,xscale=0.15]
 \begin{axis}
    [
boxplot/draw direction=y,
boxplot/variable width,
boxplot/every box/.style={fill=gray!30},
axis lines=none,
    ]
  \addplot[
    boxplot prepared={
      median=#4,
      upper quartile=#5,
      lower quartile=#3,
      upper whisker=#6,
      lower whisker=#2
    },
    ] coordinates {};
 \end{axis}
\end{tikzpicture}

}

\begin{document}
\doi{10.1080/17445760.YYYY.CATSid}
 \issn{1744-5779}
\issnp{1744-5760}
\jvol{00} \jnum{00} \jyear{2011} 

\markboth{Mohcine Chraibi and David Haensel}{Parallel, Emergent and Distributed Systems}


\title{A Knowledge-based Way-Finding Framework for Pedestrian Dynamic}

\author{Mohcine Chraibi$^{\rm a}$$^{\ast}$ \& David Haensel$^{\rm a}$\thanks{$^\ast$Corresponding author. Email: m.chraibi@fz-juelich.de
\vspace{6pt}} \\\vspace{6pt} $^{\rm a}${\em{J\"ulich Supercomputing Centre\\ Research Centre J\"ulich, J\"ulich, Germany}}\\\vspace{6pt}\received{v3.6 released March 2011} }

\maketitle

\begin{abstract}
We introduce a framework to navigate agents in buildings, inspired by the concept
of ``the cognitive map''. 
It allows to route agents depending on their spacial knowledge. 
With help of an event-driven mechanism, agents acquire new information about their surroundings, 
which expands their individual cognitive map.
\end{abstract}

\bigskip

\section{Introduction}
The simulation of pedestrian dynamics provides important results for different applications.
For architects the analysis of people flow is interesting during the planning of new facilities and exit routes.
For organizers of large scale events a simulation of pedestrians could help to appraise the location.

Hoogendoorn et al~\cite{Hoogendoorn2002} divided pedestrian dynamics decision making into three levels, the strategic, tactical and operational level.
The pre trip route planning and the choice of the final destination is done in the strategical level.
It should be mentioned that at the strategical level no information about actual circumstances is available.
Short term decisions like obstacle avoidance or route changes depending on actual situation are done at the tactical level.
At this level additional information is available, like people flow or smoked rooms.
At the operational level the pedestrian motion is modeled including interaction with other pedestrians.

Current routing mechanisms in pedestrian dynamics simulations are mostly based on shortest path calculation or quickest path approximation.
Some of them already feature perception of congestion and jams in front of doors~\cite{Kemloh2013}.
This perception leads to another route choice for some individual agents.
But most of the routing implementations do not take individual knowledge or behavior into consideration.
It is for example unrealistic to assume that pedestrians in a shopping mall take the shortest path only because they are 
knowledgeable about the exits in building.
In contrast one should assume that most pedestrians do not even know more than one emergency exit.
To resign individual knowledge assumes that every agent has perfect knowledge of the actual building.

To reach more realistic simulations it seems necessary to take some individual factors into consideration.
Individual knowledge is the basis for those individuality.
It is needed for individual decision making and social behavior.

Another important feature, the human perception, is often missing in actual implementation.
If we want to consider dynamic circumstances which influence the route choice we need to have a perception layer.
The information gathered from this perception layer should then be written in the aforementioned knowledge representation.
This shows, that a versatile knowledge representation combined with perception possibilities and decision making is needed.

\section{Related work}
\subsubsection*{Cognitive map}
The cognitive map is a concept introduced and analyzed by E.C. Tolman~\cite{Tolman1948}.
From his experiments with rats he deduced that rats are not simply navigating by stimuli and response but rather discover the space and store their acquired knowledge in a structured way.
This so called \textit{cognitive map} enables rats to make decisions while navigating.

\begin{figure}[h!]
\begin{center}
\resizebox{0.7\linewidth}{!}{
\begin{tikzpicture}[
  node distance=6cm,
  box/.style={draw, rounded corners, minimum width=4.8cm, minimum height=1cm, very thick},
  box_gray/.style={draw, rounded corners, minimum width=4.8cm, minimum height=1cm},
  ]

  \node[box_gray] (obs) at (0,0) {Observations};
  \node[box_gray] (fixed) at (-2.5, -2) {Fixed features};
  \node[box_gray] (routes) at (2.5,-3) {Routes};

  \node[box] (metric) at (-2.5, -5) {Topological map};
  \node[box] (topological) at (2.5,-6) {Metric map};

  \node[rectangle, draw, line width=1pt,  minimum width=0.5cm, minimum height=0.5cm] at (0,-4) (box) {~};
  \draw[line width=1pt,-latex] (obs) -- (fixed);
  \draw[line width=1pt,-latex] (obs) -- (routes);

  \draw[line width=1pt,-latex] (metric) |- (topological);
  \draw[line width=1pt,-latex] (fixed) -- (box);
  \draw[line width=1pt,-latex] (routes) -- (box);

  \draw[line width=1pt,-latex] (box) -| (metric);
  \draw[line width=1pt,-latex] (box) -| (topological);

\end{tikzpicture}

}
\caption[Cognitive map]{
Five different types of information of a cognitive map according to Kuipers~\cite[p. 11]{Kuipers1983}
}
\label{fig:cognitive_map_kuipers}
\end{center}
\end{figure}
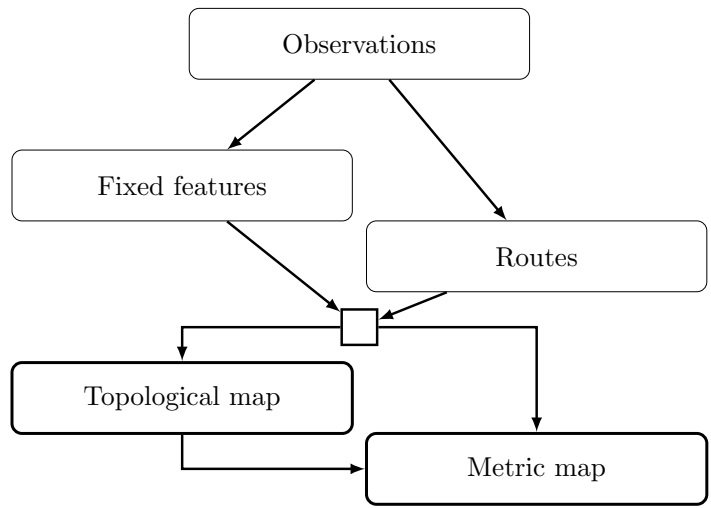

B. Kuipers later analyzed the cognitive map from a more technical point of view~\cite{Kuipers1978} and~\cite{Kuipers1983}.
In his work he described that the cognitive map aggregates information from observations to route description and fixed features which later are integrated in topological and the metric relations.
An overview of the interaction between those five types of information can be seen in figure~\ref{fig:cognitive_map_kuipers} from~\cite[p. 11]{Kuipers1983}.

\subsubsection*{Individual behavior}
Individual behavior in pedestrian dynamics simulation is handled in different ways.
Braun et al implemented individual behavior in the operational level by introducing new parameters like dependence (need of help) and altruism (willingness to help) and by introducing groups (like families)~\cite{Braun2003}.
Those new parameters change the force calculation in the underlying social force model by Helbing~\cite{Helbing2000}.
Pelechano et al used the operational level too but also took individual knowledge of the building into consideration in the way finding~\cite{Pelechano2005}.
Pan implemented a modular framework for human and social behavior which features typical behavior like queuing and leader following~\cite{Pan06}.
This new framework is also analyzed and compared with other evacuation simulations.

On the tactical level several works about way finding were performed.
The book of Arthur and Passini~\cite[chap. 5.]{Arthur1992} gives a good overview of the way finding in general and especially the process of finding a specific way.
In~\cite{Benthorn1996} the route choice during a fire is discussed.
They pay attention on the decision process and the way people choose the emergency exit (for example closed or open doors).
In addition they discuss the influence of evacuation signs and the delay time after fire alarm.

Modeling human behavior, perception and cognition  is a complicated task.
It is  not the goal of this work to reach realistic human behavior or even to understand human cognition, but to emulate the behavior simplified enough to reach adequate simulations.
With the created framework a powerful and extensible set of tools was build to model and emulate realistic behavior.

\begin{figure}[h!]

\begin{center}
\resizebox{0.7\linewidth}{!}{
\centering

  \begin{tikzpicture}[
    node distance=4cm,
    box/.style={fill=gray, text=white, rounded corners, minimum height=1cm},
    ]

      \node[box] (Perception) {Perception};
      \node[box, right of=Perception] (Knowledge) {Knowledge};
      \node[box, right of=Knowledge] (Decision) {Decisions};
      \node[rectangle, below=0.5cm of Perception, fill=gray!30, text=black, minimum height=3cm,  align=center](sensor) {Sensors \& \\  SensorManager};
      \node[cylinder, cylinder uses custom fill, cylinder body fill=gray!80, cylinder end fill=gray!80, text=white,
      shape border rotate=90, aspect=0.25, draw, minimum width=2cm, minimum height=3cm, align=center, right of=sensor]
      (cognitive_map) {Cognitive \\ Map};
      \node[rectangle, right of=cognitive_map, fill=gray!30, text=black, minimum height=3cm, align=center, minimum width=2cm] (decision_making){Decision  \\  Making};
      \draw[decorate, decoration={snake,amplitude=.4mm,segment length=2mm,post length=1mm},  -Stealth] (sensor.20) -- (cognitive_map);
      \draw[decorate, decoration={snake,amplitude=.4mm,segment length=2mm,post length=1mm},  -Stealth] (sensor) -- (cognitive_map);
      \draw[decorate, decoration={snake,amplitude=.4mm,segment length=2mm,post length=1mm},  -Stealth] (sensor.340) -- (cognitive_map);
      \draw[line width=2pt,   -Stealth] (cognitive_map) -- (decision_making);

  \end{tikzpicture}

}

\caption[Modules interaction overview]{Overview of the interaction between the modules.}

\label{overview_cm}
\end{center}
\end{figure}
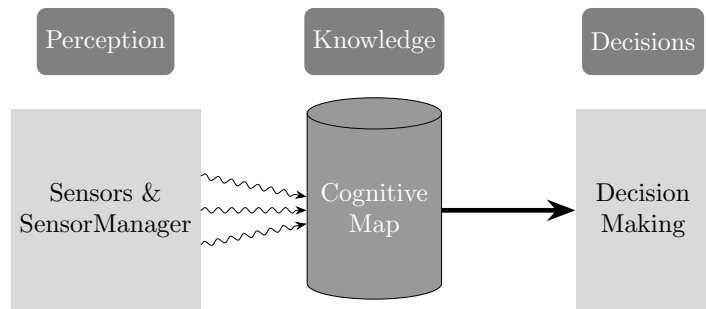

For achieving individual behavior and basic reasoning in pedestrian dynamics simulations it is necessary to have a versatile spatial knowledge representation.
For this reason a simplified version of the cognitive map  proposed by Tolman~\cite{Tolman1948} was implemented and used for each agent separately (section~\ref{simplified-cm}). 
To model the information gathering of pedestrians a sensor structure was build to enrich the information stored in the cognitive map (section \ref{information-collection}).
Moreover the stored knowledge is used for individual decision making (section \ref{decision-making}).
The aforementioned three modules are encapsulated in the new created cognitive map router.
Figure \ref{overview_cm} shows an overview of the build modules which are described in detail in the following sections.
\section{Representing knowledge with a cognitive map}
\label{simplified-cm}
The knowledge representation is the central module for the new routing mechanism.
It has to fulfil several requirements.
One of them is the possibility to represent different spatial knowledge.
Another one is the possibility to store knowledge individually to achieve the goal of individual behaviour.

For the data structure we use the concept of the cognitive map proposed by Tolman \cite{Tolman1948} which is widely used in robotics navigation. 
Due to the fact that the navigation is used for agents in evacuation simulations inside a building we omit some parts of the cognitive map concept.
A metric map (section \ref{CM_navigation_graph}) and a memory of used routes (section \ref{CM_used_routes}) constitute our new cognitive map (section \ref{CM_cm}).
The metric map represents the notion which the simulated pedestrian has about the building whereas the memory of used routes constitutes a simple remembrance.

\subsection{Requirements and preconditions}
\label{CM_preconditions}
\begin{figure}[h!]
\begin{center}
\resizebox{0.7\linewidth}{!}{
\begin{tikzpicture}[
  wall/.style={line width=2pt},
  door/.style={line width=2pt, color=gray!30},
  exit/.style={line width=2pt, color=black!70, densely dotted},
  white/.style={line width=3pt, color=white},
]

  \draw[wall] (0,0) -- (0,8) -- (14,8) -- (14,0) -- cycle;
  \draw[wall] (0,4) -- (6.5,4) -- (6.5,8);
  \draw[wall] (14,4) -- (7.5,4) -- (7.5,8);
  \draw[wall] (0,3) -- (14,3);
  \draw[wall] (5,3) -- (5,0);
  \draw[wall] (12,3) -- (12,0);

  \draw[white] (0,3) -- (0,4);
  \draw[white] (14,3) -- (14,4);
  \draw[white] (8,0) -- (9,0);
  \draw[exit] (0,3) -- (0,4);
  \draw[exit] (14,3) -- (14,4);
  \draw[exit] (8,0) -- (9,0);
  \draw[door] (2,3) -- (3,3);
  \draw[door] (6.5,3) -- (7.5,3);
  \draw[door] (10,3) -- (11,3);
  \draw[door] (2,4) -- (3,4);
  \draw[door] (12,1.25) -- (12,1.75);
  \draw[door] (10,4) -- (11,4);
  \draw[door] (6.5,5.5) -- (6.5,6.5);

  \draw[door] (6.55,4) -- (7.45,4);

\node[] at (7, -0.7) {\fbox{
\begin{tabular}{rlrlrlrl}
 \raisebox{2pt}{\tikz{\draw[wall] (0,0) -- (5mm,0);}}&Wall&
 \raisebox{2pt}{\tikz{\draw[door] (0,0) -- (5mm,0);}}&Door&
 \raisebox{2pt}{\tikz{\draw[exit] (0,0) -- (5mm,0);}}&Exit&

\end{tabular}}};

\end{tikzpicture}
}
\caption[Floor plan example]{An example of a floor plan used for simulations.}
\label{fig:floor_plan}
\end{center}
\end{figure}
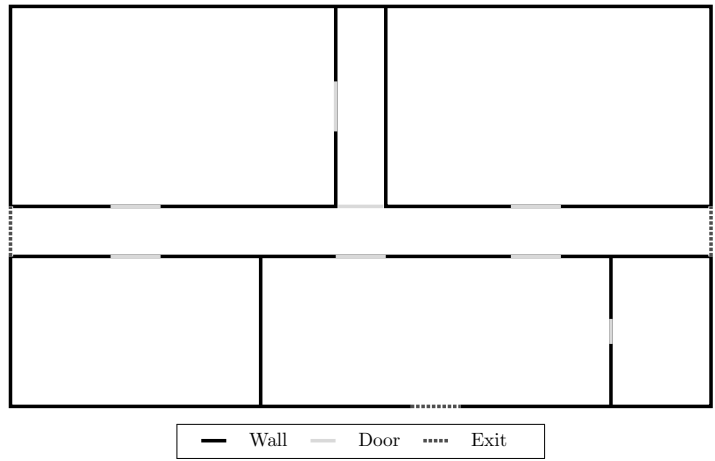
Figure \ref{fig:floor_plan} shows a simple floor plan with five rooms and two corridors.
In the simulation software those floor plans have to be converted to another structure.
This structure is important.

The simulation software divides the building in several rooms which are divided in further sub rooms.
The rooms and sub rooms are bounded by walls, crossings and transitions (for example doors).
A crossing is the intersection between two sub rooms among the same room and a transition is the intersection between two different rooms or a room and the outside.


From the geometry structure we deduced the possible spatial knowledge and classified it into two classes.
The basic structure an agent could be aware of are sub rooms and doors.
The knowledge of existence of sub rooms and doors are then classified as first order knowledge.
The second order knowledge are additional information an agent could have about sub rooms or doors (e.g. the blockage of a door or the smoke emergence in a sub room).
With this classification second order knowledge is always related to elements of the first order.
For example the awareness of a congestion in front of a door is knowledge of second order whereas the knowledge of the door itself belongs to the first order knowledge.
Another advantage is the possibility to store knowledge of second order as a property of an element of first order.
Therefor the modeling of the new data structure has to consider the first order knowledge primarily.
In table \ref{tab:kowledge_classification} some illustrative examples are shown.

\begin{table}[h]
\begin{center}
  \begin{tabular}{ r p{9cm} }
      \hline
      First order knowledge item & Corresponding second order knowledge examples \\
      \hline
      Sub room & Smoke in the sub room. \\
           & Type of the sub room (e.g. corridor or normal sub room). \\
      \hline
      Door & The people density in front of the door. \\
           & A congestion at the door.  \\
           & The blockage of a door. \\
      \hline
  \end{tabular}
  \caption[Knowledge classification examples]{Examples for first and second order spatial knowledge.}
  \label{tab:kowledge_classification}
\end{center}
\end{table}

Even if our first order knowledge is bound to sub rooms and doors the second order knowledge is really versatile.
We just mentioned some examples but a lot of other information would be imaginable.
The first order knowledge is a robust spatial representation whereas the second order knowledge gives us the versatility.


\subsection{The metric map (NavigationGraph)}
\label{CM_navigation_graph}
The metric map constitutes the notion of the building an agent has.
Therefore it has to represent the knowledge of first order as well as the knowledge of second order.
As aforementioned elements of second order could be modeled as properties of elements of first order.
Therefore we have to care about first order elements mostly.

For the first order knowledge we decided to use a graph based structure.
That is why we had to identify sub rooms and doors with vertices and edges.
In contrast to some former routing algorithm we identify the sub rooms as vertices and the doors as edges. 

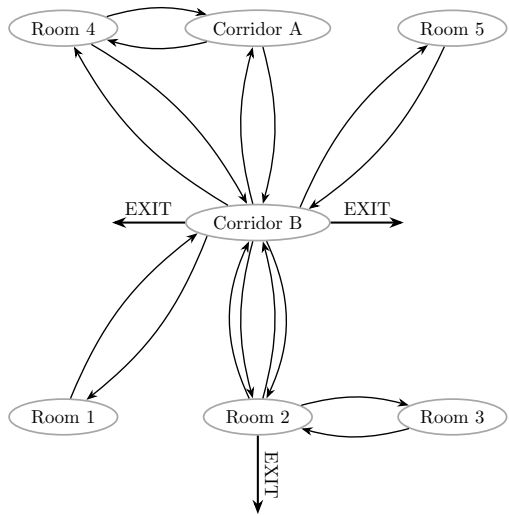
\begin{figure}[h!]
\begin{center}
\resizebox{0.5\linewidth}{!}{
\begin{tikzpicture}[
  node distance = 4cm,
  wall/.style={line width=2pt},
  door/.style={line width=2pt, color=gray!30},
  exit/.style={line width=2pt, color=gray},
  transition/.style={line width=1pt, color=gray!70, densely dotted},
  vertex/.style={line width=1pt, draw=gray!70, ellipse, rounded
    corners},
  edge/.style={-Stealth, thick},
]

  \node[vertex] (room_1) {Room 1};
  \node[vertex] (room_2) [right of=room_1] {Room 2};
  \node[vertex] (room_3) [right of=room_2] {Room 3};
  \node[vertex] (corridor_b) [above of=room_2] { Corridor B};
  \node[vertex] (corridor_a) [above of=corridor_b] { Corridor A};

  \node[vertex] (room_4) [left of=corridor_a] {Room 4};

  \node[vertex] (room_5) [right of=corridor_a] {Room 5};


  \draw [edge] (room_1) to[bend left=15] (corridor_b.190);
  \draw [edge] (corridor_b.195) to[bend left=15] (room_1);
  \draw [edge] (room_2) to[bend left=25] (corridor_b);
  \draw [edge] (corridor_b) to[bend right=15] (room_2);
  \draw [edge] (room_2) to[bend right=15] (corridor_b);
  \draw [edge] (corridor_b) to[bend left=25] (room_2);
  \draw [edge] (room_5) to[bend left=15] (corridor_b.15);
  \draw [edge] (corridor_b.20) to[bend left=15] (room_5);
  \draw [edge] (room_4) to[bend left=15] (corridor_b);
  \draw [edge] (corridor_b) to[bend left=15] (room_4);
  \draw [edge, bend left=25] (room_4) to[bend left=15]  (corridor_a);
  \draw [edge] (corridor_a) to[bend left=15] (room_4);
  \draw [edge] (corridor_a) to[bend left=15] (corridor_b);
  \draw [edge] (corridor_b) to[bend left=15] (corridor_a);
  \draw [edge] (room_2) to[bend left=15] (room_3);
  \draw [edge] (room_3) to[bend left=15] (room_2);
  \draw [edge, very thick] (corridor_b) -- node[sloped,
  above]{EXIT} +(-3,0);
  \draw [edge, very thick] (corridor_b) -- node[sloped, above]{EXIT} +(3,0);
  \draw [edge, very thick] (room_2) -- node[sloped, above]{EXIT}  +(0,-2);

\end{tikzpicture}

}
\caption[Graph from floor plan]{The complete navigation graph generated from the floor plan of figure \ref{fig:floor_plan}.}
\label{fig:graph}
\end{center}
\end{figure}

This is reasonable in order to have a versatile structure for adding information to a certain edge.
In our representation an edge represents the intersection between sub rooms, which is needed to guide agents from room to room instead of guiding them from door to door.
Another advantage is the possibility to store different information for different edges directions.
For example leaving a room towards a corridor is rated better than the other direction.
This structure gives us the possibility to have an idea about leaving the sub room in the first place, which would be difficult if doors are vertices.
The chosen structure has some downsides as well.
When it comes to accurate distance calculations some problems appear.
It is possible to create a sparse graph to model pedestrians with incomplete knowledge of a building.
Figure \ref{fig:graph} shows the finished graph used inside the simulation software after processing the geometry.
In this example the navigation graph contains all sub rooms and doors thus it is complete.

\clearpage
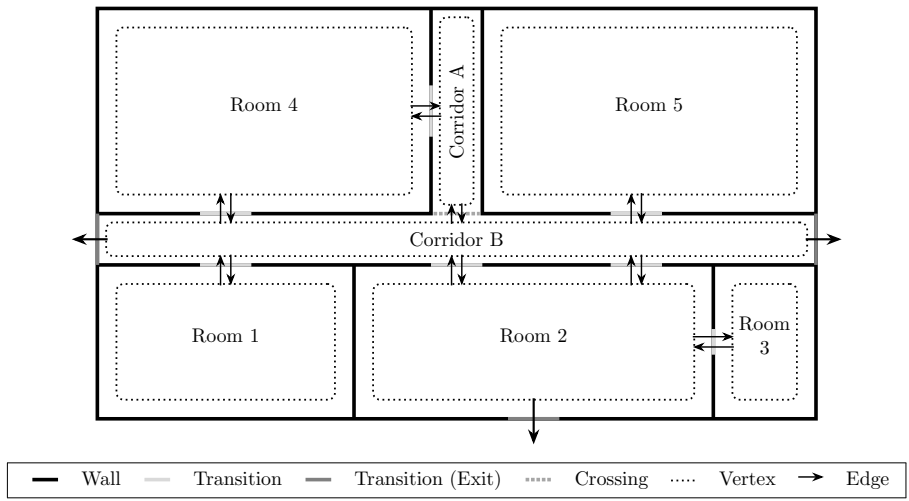
\begin{figure}[ht]
\begin{center}
\resizebox{0.9\linewidth}{!}{
\begin{tikzpicture}[
  wall/.style={line width=2pt},
  door/.style={line width=2pt, color=gray!30},
  exit/.style={line width=2pt, color=gray},
  transition/.style={line width=2pt, color=gray!70, densely dotted},
  vertex/.style={line width=1pt, draw=black, rectangle, rounded
    corners, dotted},
  edge/.style={-Stealth, thick},
]

  \draw[wall] (0,0) -- (0,8) -- (14,8) -- (14,0) -- cycle;
  \draw[wall] (0,4) -- (6.5,4) -- (6.5,8);
  \draw[wall] (14,4) -- (7.5,4) -- (7.5,8);
  \draw[wall] (0,3) -- (14,3);
  \draw[wall] (5,3) -- (5,0);
  \draw[wall] (12,3) -- (12,0);

  \draw[exit] (0,3) -- (0,4);
  \draw[exit] (14,3) -- (14,4);
  \draw[exit] (8,0) -- (9,0);
  \draw[door] (2,3) -- (3,3);
  \draw[door] (6.5,3) -- (7.5,3);
  \draw[door] (10,3) -- (11,3);
  \draw[door] (2,4) -- (3,4);
  \draw[door] (12,1.25) -- (12,1.75);
  \draw[door] (10,4) -- (11,4);
  \draw[door] (6.5,5.5) -- (6.5,6.5);

  \draw[transition] (6.55,4) -- (7.45,4);

  \node[vertex, fit={(0.5,0.5) (0.5,2.5) (4.5,2.5) (4.5,0.5)}]
  (room_1) {Room 1};
  \node[vertex, fit={(5.5, 0.5) (5.5,2.5) (11.5,2.5) (11.5,0.5)}]
  (room_2) {Room 2};
  \node[vertex, fit={(12.5,0.5) (12.5,2.5) (13.5,2.5) (13.5,0.5)}]
  (room_3) {Room 3};
  \node[vertex, fit={(0.5,4.5) (0.5,7.5) (6,7.5) (6,4.5)}] (room_4)
  {Room 4};
  \node[vertex, fit={(8,4.5) (8,7.5) (13.5,7.5) (13.5,4.5)}] (room_5)
  {Room 5};

  \node[vertex, fit={(6.8,4.3) (6.8,7.7) (7.2,7.7) (7.2,4.3)}]
  (corridor_a) { \\ \rotatebox{90}{Corridor A}};
  \node[vertex, fit={(0.3,3.3) (0.3,3.7) (13.7,3.7) (13.7,3.3)}]
  (corridor_b) { \\[-5pt] Corridor B };

  \draw [edge] (2.4, 2.6) -- (2.4, 3.2);
  \draw [edge] (2.6, 3.2) -- (2.6, 2.6);
  \draw [edge] (6.9, 2.6) -- (6.9, 3.2);
  \draw [edge] (7.1, 3.2) -- (7.1, 2.6);
  \draw [edge] (10.4, 2.6) -- (10.4, 3.2);
  \draw [edge] (10.6, 3.2) -- (10.6, 2.6);
  \draw [edge] (10.4, 3.8) -- (10.4, 4.4);
  \draw [edge] (10.6, 4.4) -- (10.6, 3.8);
  \draw [edge] (2.4, 3.8) -- (2.4, 4.4);
  \draw [edge] (2.6, 4.4) -- (2.6, 3.8);
  \draw [edge] (6.1, 6.1) -- (6.7, 6.1);
  \draw [edge] (6.7, 5.9) -- (6.1, 5.9);
  \draw [edge] (6.9, 3.8) -- (6.9, 4.2);
  \draw [edge] (7.1, 4.2) -- (7.1, 3.8);
  \draw [edge] (11.6, 1.6) -- (12.4, 1.6);
  \draw [edge] (12.4, 1.4) -- (11.6, 1.4);
  \draw [edge, very thick] (0.2, 3.5) -- (-0.5, 3.5);
  \draw [edge, very thick] (13.8, 3.5) -- (14.5, 3.5);
  \draw [edge, very thick] (8.5, 0.4) -- (8.5, -0.5);

\node[] at (7, -1.2) {\fbox{
\begin{tabular}{rlrlrlrlrlrl}
 \raisebox{2pt}{\tikz{\draw[wall] (0,0) -- (5mm,0);}}&Wall&
 \raisebox{2pt}{\tikz{\draw[door] (0,0) -- (5mm,0);}}&Transition&
 \raisebox{2pt}{\tikz{\draw[exit] (0,0) -- (5mm,0);}}&Transition (Exit)&
 \raisebox{2pt}{\tikz{\draw[transition] (0,0) -- (5mm,0);}}&Crossing&
 \raisebox{2pt}{\tikz{\draw[vertex] (0,0) -- (5mm,0);}}&Vertex&
 \raisebox{2pt}{\tikz{\draw[edge] (0,0) -- (5mm,0);}}&Edge
\end{tabular}}};

\end{tikzpicture}

}
\caption[Floor plan with graph]{The floor plan used in simulations with assigned complete navigation graph structure.}
\label{fig:graph_floor_plan}
\end{center}
\end{figure}

\subsubsection{Implementation}
The implementation of the navigation graph consists out of three classes the \cclass{NavigationGraph}, the \cclass{Vertex} and the \cclass{Edge} class.
The \cclass{NavigationGraph} class exists out of a collection of pointers to \cclass{Vertex} objects and several method to manipulate or read parts of the graph.
The \cclass{Vertex} class consists out of a pointer to the corresponding \cclass{SubRoom} and a collection of \cclass{Edges} starting in this vertex.
The \cclass{Edge} class has a source and a destination pointer to the corresponding \cclass{Vertex} and a pointer to the \cclass{Crossing} or \cclass{Transition}.
The \cclass{Edge} has a collection of factors which are used to calculate the weight of an edge.
This weight is later used for decision making.



\subsubsection{Edge-weight calculation}
The calculation of optimal routes is based on the weight of each edge.
These depend primarily on the distance and other factors which represent the second order knowledge.

Let $e_i \in E$ be an edge of the navigation graph $G = G(V,E)$.
$$F_i \coloneqq \{ f_k \in F_i \  | \  f_k \in \mathbb{R} \wedge  f_k > 0\}\qquad \qquad (|F_i| < \infty) $$ 

is the \textbf{set of corresponding factors} for $e_i$. The elements of $F_i$ are called \textbf{edge-factors}.

$$f^ {(i)} \coloneqq \prod_{f_k \in F_i} f_k$$ is called the \textbf{accumulated edge-factor}.

Let $x_i$ be the length of the edge $e_i$ and $F_i$ the set of corresponding factors for $e_i$. Then is
\begin{equation}
w_i \coloneqq x_i \cdot \prod_{f_k \in F_i} f_k = x_i \cdot f^{(i)}
\end{equation}
the \textbf{edge-weight} of $e_i$.

Edge-factors and sensors are highly related to the decision making process.
The decision making is based on the edge-weight, to decide for an optimal route.
Thus the edge-factors have a high influence on the chosen route.

\subsection{Used routes memory}
\label{CM_used_routes}
The smaller part of the cognitive map is the memory of used routes, which is for the purpose of representing the pedestrians remembrance.
We store every edge which was chosen by the decision making in the same order.
With this we can reconstruct the chosen path.
An application could be a sensors which avoids the agent from going backwards.

\subsection{Putting it all together: the cognitive map}
\label{CM_cm}
Even if this cognitive map is a drastic simplification of Tolmans cognitive map it is less complex and still fits our needs.

In the simulation each agent has its individual cognitive map.
These maps are accessed through the \textit{CognitiveMapStorage} class which also takes care of the creation of the initial cognitive maps.
The creation itself is done by \textit{CognitiveMapCreator} classes which are passed to the \textit{CognitiveMapStorage} and executed when needed.
With these \textit{CognitiveMapCreators} it is possible to create different cognitive maps, in matter of information content, to simulate pedestrians with different knowledge.
It is also possible to use different creators for different agents.
The current implementation features two creators the \textit{CompleteCognitiveMapCreator} which creates a complete cognitive map and an \textit{EmptyCognitiveMapCreator} which creates an empty cognitive map.
Further creators can be easily implemented.

\subsection{Gathering information}
\label{information-collection}
With the proposed cognitive map we have a versatile structure for the later discussed decision making.
The decision making is based on the edge weight and thereby on the edge factors.
Therefore it is important that the cognitive map in general and the edge factors in particular are up to date to make current decisions.

The information gathering module is responsible for this update process.
It provides a framework for reproducing simplified human perception.
It is able to manipulate edge factors as well as an entire edge or vertex.

For gathering information a sensor structure was build.
These sensors are managed and executed by an event driven sensor manager.
Those events could be triggered in every time step during the execution of the router.
Therefore the sensors are executed individually for each agent.

\subsection{Sensors and sensor manager}
\label{information-collection-sensors}
The sensor system consists of \cclass{Sensors} and the \cclass{SensorManager} class.

\subsubsection{SensorManager and process}
The sensor system was build with the observer design pattern in mind.
The \cclass{SensorManager} is the observing object which observes the \cclass{CognitiveMapRouter} object.
An object inherited from \cclass{AbstractSensor} could be registered to the \cclass{SensorManager} for certain events.
When the \cclass{CognitiveMapRouter} reaches an event state it notifies the \cclass{SensorManager}, which then executes the \cclass{Sensors} registered for this event.


Table \ref{tab:events} shows the available events.
The integration of further events is possible.
\begin{table}[h!]
\centering
  \begin{tabular}{ r p{11cm} }
      \hline
      Event name & Description \\
      \hline
      \cclass{INIT} & This event is triggered during the initialization of the cognitive map of the respective agent. \\
      \cclass{CHANGED ROOM} &  This event is triggered when an agent changed the sub room. \\
      \cclass{NEW DESTINATION} & This event is triggered after the agent got a new destination from the router. \\
      \cclass{NO WAY} & This event is triggered when the agent could not find a way to a known emergency exit. \\
      \hline
  \end{tabular}
 \caption[Available events] {Available events for registering \cclass{Sensors} to the \cclass{SensorManager}.}
 \label{tab:events}
\end{table}

With the \cclass{SensorManager} it is possible to register \cclass{Sensors} as needed.
So its the task of the user to decide which sensors should be activated and used on which event.
It is not even required to use one of them.
With this, the sensor execution chain is completely adjustable to the needs of the actual simulation.

\subsubsection{Sensors}
A \cclass{Sensor} is a class inherited from \cclass{AbstractSensor}.
The task of the sensor is the manipulation and creation of first and second order knowledge of the cognitive map.
Therefor it could read from the \cclass{CognitiveMap}, the \cclass{Building} and the corresponding \cclass{Pedestrian} object.
The \cclass{Sensor} could be thought as both, a modeled sensing device or a information manipulator without any physical sensing in mind.
It is the main input for the cognitive map.
It could add or delete edges and vertices or manipulate edge factors.

With the versatile sensors it is possible to add any information to the cognitive map and thereby change the decision making.
For example one could add data from a fire simulation to have a better smoke status of a room.
This smoke data could then be used to set a defined edge factor to edges heading to the smoked sub room.
This will lead to a decision in favour of non smoked sub rooms.
So the sensor could be the interface for getting more data into the routing and decision making process.

\subsubsection{Implementing a sensor}
A new sensor has to inherit from the abstract class \cclass{AbstractSensor}.
The parent class specifies the implementation of an execute function and a get name function.
Furthermore a constructor function is inherited  which sets a pointer to the \cclass{Building} object.
The \cclass{GetName} function is mostly used for storing factors and sensors identified by name.
The \cclass{execute} function is the main function of the sensor.
This function is executed when reaching an event for which the sensor is registered.
In this function the \cclass{CognitiveMap}, \cclass{Building} and \cclass{Pedestrian} object could be used to manipulate the \cclass{CognitiveMap}.

An implemented sensor could than be registered to the \cclass{SensorManager}.

\subsection{Implemented sensors}
\label{implemented_sensors}

In this section the implemented sensors are presented.

\subsubsection{RoomToCorridorSensor}
The \cclass{RoomToCorridorSensor} adds an edge-factor to the respective edge depending on the source and destination sub room type.
It cares about sub rooms of the type room or corridor.
Based on the assumption, that changing a room in direction of an exit corridor is good and leaving an exit corridor in the direction of a usual room is bad, the edge-factor is lower or higher than one.
This edge-factor makes agents tend to go in the direction of a corridor or to stay on corridors. 

\subsubsection{LastDestinationsSensor}
The \cclass{LastDestinationsSensor} is based on the used routes memory.
It sets a penalty edge-factor ($>1$) to the corresponding edge in opposite direction if it exists.
This way an agent is hindered from going back immediately.
However raising the edge factor only means that the chance of going back is minimal but not zero, since the edge still exists in the navigation graph.
The sensor sets the edge-factor which raises the weight but does not delete any edge.
If an agent has discovered all possible directions he could decide to go back again if all other edges have even worse weights (for more details on decision making see section \ref{decision-making}).

\subsubsection{DiscoverDoorsSensor}
The \cclass{DiscoverDoorsSensor} should be used after the agent was not able to find any route to an emergency exit.
It emulates the process of discovering an unknown room.
This sensor adds all possible out edges to the vertex corresponding to the actual sub room.
After the sensor execution the agent can at least start searching for a local route.

\subsubsection{SmokeSensor}
The \cclass{SmokeSensor} sets a smoke edge-factor.
This edge-factor should consider the smoke status or fire hazard in a certain sub room.
If the smoke status of a sub room is activated, the sensor raises the smoke edge-factor of all edges heading to this sub room.
In this way the agent is encouraged to avoid the smoked sub room.

\subsubsection{DensitySensor}
The \cclass{DensitySensor} measures the density in front of a door (crossing or transition) and sets the corresponding edge-factor.
If the density is too high the agent tends to take another route and avoid the jam.

\subsection{Making decisions}
\label{decision-making}
The last module of the proposed routing framework is the decisions making module.
With the already defined edge-factors and the deduced edge-weight it was nearby to calculate optimal routes in the given navigation graph.
But due to the fact, that some agents may have a sparse navigation graph it is possible  that an agent does not know any complete exit route.
There for we distinguish between agents with enough knowledge to find a complete exit route and agents with less knowledge.
The first group uses a global optimization algorithm and the second uses a local algorithm.
Till now there is a strong separation of strategies of this two groups, but for more realistic behavior a combination of both strategies would be advisable.

\section{Simulation results}
To showcase the flexibility of our new routing framework we analyze the impact of different configurations on the route choice and thus on the whole simulation.
The implemented routing framework is adjustable in many different ways.
This leads to a lot of possible configurations, which we can not analyze in all aspects.
That is why we choose some configurations for the analysis.
As first we show the behavior of a pedestrian with an empty cognitive map and how it manages to leave the building. 
Second,  we investigate the impact of the density sensor when used with complete cognitive maps.

\subsection{Mandatory sensors only}
\label{sec:sim_NE}
As mentioned before the \cclass{LastDestinationsSensor} and the \cclass{DiscoverDoorsSensor} are mandatory for simulations with empty cognitive maps.
With those two sensors and the \cclass{EmptyCognitiveMapCreator} the agent knows nothing at first.
After the first run of the \cclass{DiscoverDoorsSensor} the agent knows at least the doors of the actual sub room.
Since the edge factors do not differ, at least if the agent did not pass the door already (\cclass{LastDestinationsSensor}), the agent always chooses the door with minimal distance which extends the average evacuation path length drastically.

This reveals some simulations where the agent explores every door before the corridor with the emergency exits is found.
This is not realistic but caused by the absence of any sensor and thus any valuable information which could provide something like a heuristic.
In real life situations humans would rarely evacuate without any heuristic.
For demonstration purposes we conducted simulations without additional sensors anyway to show the change in evacuation time with and without additional sensors.

\begin{figure}[H]
\begin{center}
  \includegraphics[width=0.8\linewidth]{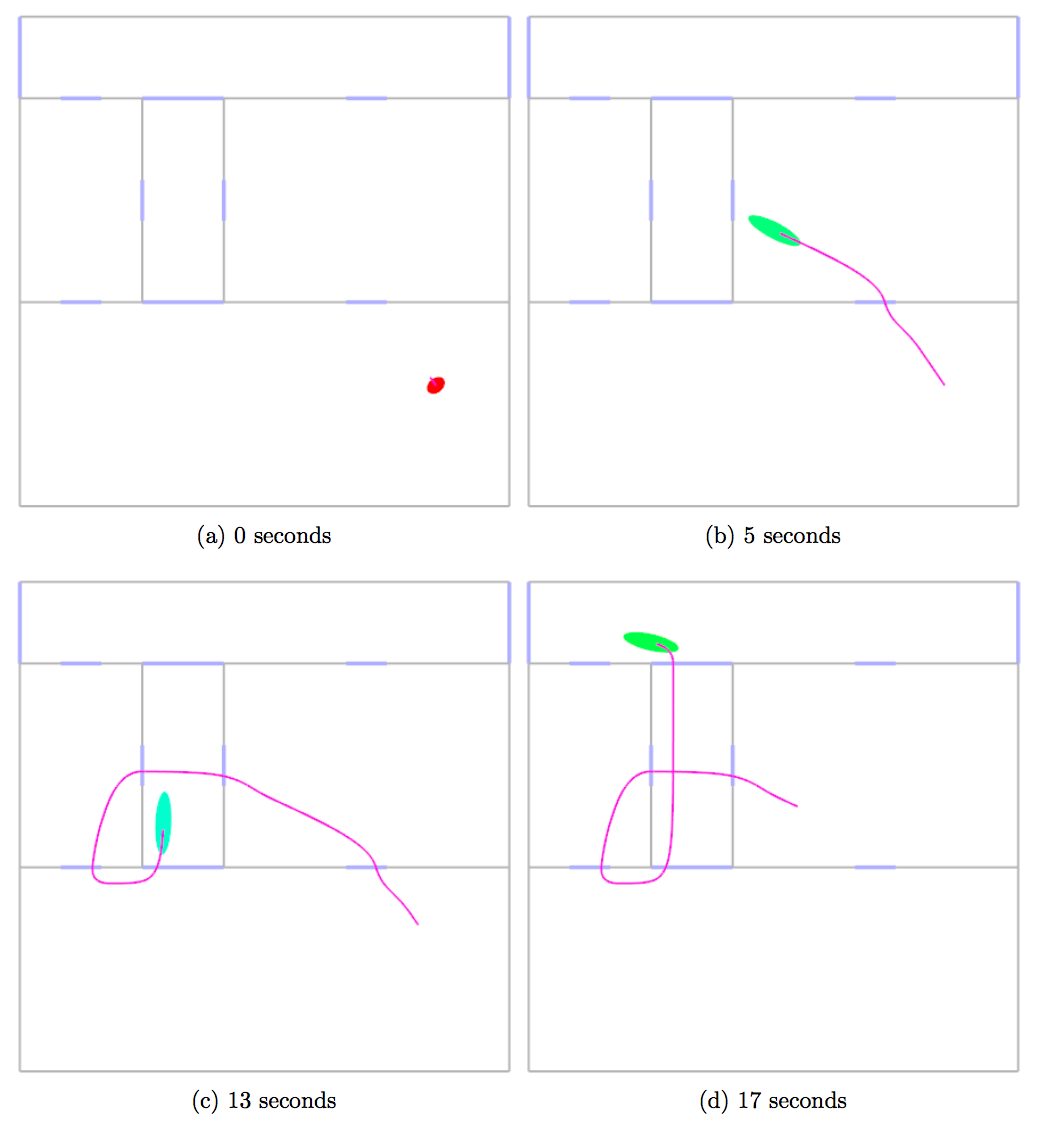}

        \caption[Simulation with empty cognitive map]{
          Simulation with empty cognitive map and no additional sensors.
        }
        \label{sim:EmptyNone}
\end{center}
\end{figure}

Table \ref{tab:sim_empty_none} shows the evacuation time frequencies of one agent in 50 simulations.
Figure \ref{sim:EmptyNone} shows the described behavior with some snapshots from one simulation.


\begin{table}[h]

\centering
\begin{tabular}{ l   l  l l l}
  \toprule
  \multicolumn{1}{l}{\textbf{t [s]}} & \multicolumn{1}{l}{\textbf{Frequency}} & \multicolumn{2}{c}{\textbf{Percentage}} \\[5pt]
  \toprule

  18 & 1  & \percentright{0.02} &\multirow{9}{*}[14pt]{\boxplot[1.15]{18}{20}{22}{23}{26}} \\
  19 & 6  & \percentright{0.12} & \\
  20 & 11 & \percentright{0.22} & \\
  21 & 4  & \percentright{0.08} & \\
  22 & 10 & \percentright{0.20} & \\
  23 & 8  & \percentright{0.16} & \\
  24 & 8  & \percentright{0.16} & \\
  25 & 1  & \percentright{0.02} & \\
  26 & 1  & \percentright{0.02} & \\[2pt]

  \bottomrule
\end{tabular}
\caption[Empty cognitive map no additional sensors]{
  Evacuation time frequencies of 50 simulations with \cclass{EmptyCognitiveMapCreator} and no additional sensors.
  Simulations are done with one agent and \cclass{LastDestinationsSensor} and \cclass{DiscoverDoorsSensor}.
}
\label{tab:sim_empty_none}
\end{table}

\subsection{Density sensor}

For the analysis we conduct 50 simulations for each configuration of interest with different initial conditions.
The initial conditions influence mainly the position of pedestrians in a certain sub room.

The initial conditions are different among the 50 simulations of one configuration but equal for different configurations.
With this setup we want to minimize the influence of random effects.
For some configurations we compare the total evacuation time with the total evacuation time obtained using the global shortest path router~\cite{Kemloh2013}.
The total evacuation time is the duration until the last agent has left the building.
For comparing evacuation times we use Welch's t-test \cite{Welch1947}.
This test has the null hypothesis that the expected values of the distributions of the two samples are the same using the mean value.
Rejecting this null hypothesis means that the expected values are significantly different.
All tests are calculated with R's \cite{RLang} intrinsic t-test method.
All agents have a complete initial cognitive map and thus ``know'' every sub room and door in the building.
The only additional information which could be added is knowledge of second order.

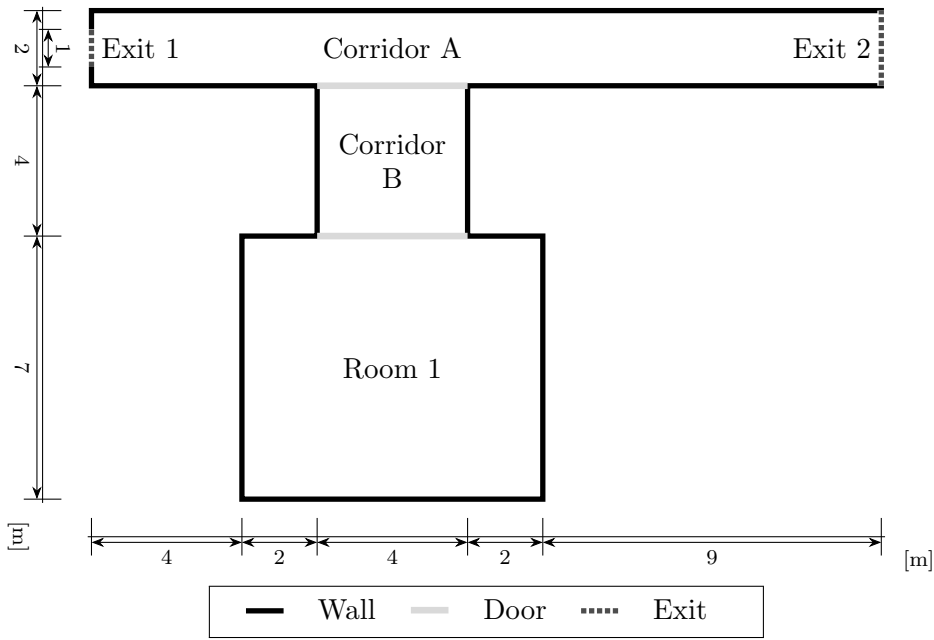
\begin{figure}[h!]
\begin{center}
\begin{tikzpicture}[
  scale=0.5,
  wall/.style={line width=2pt},
  door/.style={line width=2.5pt, color=gray!30},
  exit/.style={line width=2pt, color=black!70, densely dotted},
  white/.style={line width=3pt, color=white},
  measure/.style={node font=\footnotesize},
  grid/.style={color=gray!50, dashed},
]

  \draw[wall] (0,13) -- (21,13) -- (21,11) -- (10,11) -- (10,7) -- (12,7) -- (12,0) -- (4,0) -- (4,7) -- (6,7) -- (6,11) -- (0,11) -- cycle;

  \draw[white] (0,11.5) -- (0,12.5);
  \draw[white] (21,11) -- (21,13);

  \draw[exit] (0,11.5) -- (0,12.5);
  \draw[exit] (21,11) -- (21,13);

  \draw[door] (6,7) -- (10,7);

  \draw[door] (6,11) -- (10,11);

  \node[align=center]  at (8,3.5) {Room 1};

  \node[align=center]  at (8,12) {Corridor A};
  \node[align=center]  at (8,9) {Corridor \\ B};

  \node[align=center, anchor=west]  at (0,12) {Exit 1};
  \node[align=center, anchor=east]  at (21,12) {Exit 2};

  \draw [] (-0.1,-1) -- (21.1,-1);
  \draw [] (0,-1.5) -- (0,-0.5);
  \draw [] (4,-1.5) -- (4,-0.5);
  \draw [] (6,-1.5) -- (6,-0.5);
  \draw [] (10,-1.5) -- (10,-0.5);
  \draw [] (12,-1.5) -- (12,-0.5);
  \draw [] (21,-1.5) -- (21,-0.5);

  \draw[Stealth-Stealth] (0,-1.15) -- (4,-1.15);
  \draw[Stealth-Stealth] (4,-1.15) -- (6,-1.15);
  \draw[Stealth-Stealth] (6,-1.15) -- (10,-1.15);
  \draw[Stealth-Stealth] (10,-1.15) -- (12,-1.15);
  \draw[Stealth-Stealth] (12,-1.15) -- (21,-1.15);

  \node[anchor=north, measure] at (22, -1.15) {[m]};
  \node[anchor=north, measure] at (2, -1.15) {4};
  \node[anchor=north, measure] at (5, -1.15) {2};
  \node[anchor=north, measure] at (8, -1.15) {4};
  \node[anchor=north, measure] at (11, -1.15) {2};
  \node[anchor=north, measure] at (16.5, -1.15) {9};

  \draw [] (-1.3,-0.1) -- (-1.3,13.1);
  \draw [] (-1.8,0) -- (-0.8,0);
  \draw [] (-1.8,7) -- (-0.8,7);
  \draw [] (-1.8,11) -- (-0.8,11);
  \draw [] (-1.8,13) -- (-0.8,13);

  \draw [] (-1.4,11.5) -- (-0.8,11.5);
  \draw [] (-1.4,12.5) -- (-0.8,12.5);

  \draw[Stealth-Stealth] (-1.45, 0) -- (-1.45, 7);
  \draw[Stealth-Stealth] (-1.45, 7) -- (-1.45, 11);
  \draw[Stealth-Stealth] (-1.45, 11) -- (-1.45, 13);
  \draw[Stealth-Stealth] (-1.15, 11.5) -- (-1.15, 12.5);

  \node[anchor=north, measure, rotate=270] at (-1.45,-1) {[m]};
  \node[anchor=north, measure, rotate=270] at (-1.45,3.5) {7};
  \node[anchor=north, measure, rotate=270] at (-1.45,9) {4};
  \node[anchor=north, measure, rotate=270] at (-1.45,12) {2};
  \node[anchor=south, measure, rotate=270] at (-1.15,12) {1};

\node[] at (10.5,-3) {\fbox{
\begin{tabular}{rlrlrlrl}
 \raisebox{2pt}{\tikz{\draw[wall] (0,0) -- (5mm,0);}}&Wall&
 \raisebox{2pt}{\tikz{\draw[door] (0,0) -- (5mm,0);}}&Door&
 \raisebox{2pt}{\tikz{\draw[exit] (0,0) -- (5mm,0);}}&Exit&

\end{tabular}}};

\end{tikzpicture}
\caption[Geometry for analysis of DensitySensor]{
  The geometry (T-Junction) used for analyzing the \textit{DensitySensor}.
}
\label{fig:floor_plan_analysis_tjunction}
\end{center}
\end{figure}

For the analysis of the \cclass{DensitySensor} we conducted simulations with the geometry shown in figure~\ref{fig:floor_plan_analysis_tjunction}.
Exit 1 is smaller than exit 2 and should cause some congestion.
In addition the way from room 1 to exit 1 is shorter than the way to exit 2.
We distributed 140 agents in room 1.
Without any sensors the agents take the direct path to the exit 1.
This leads to a high density and a congestion in front of the emergency exit.
With the \cclass{DensitySensor} the exit 2 is used too and the agents are distributed better.
Depending on the actual density when arriving at corridor A the agents decides whether to go to exit 1 or to exit 2.
The total evacuation time is significant lower with the \cclass{DensitySensor} than without.
Table \ref{tab:sim_tjunction_complete_Density} shows the comparison of the total evacuation times.



\begin{table}[H]
\setlength\extrarowheight{3pt}
\centering
\begin{tabular}{ l |   r  r r || l l l}
  \hline
  \multicolumn{1}{l}{\textbf{t [s]}} & \multicolumn{1}{l}{\textbf{Freq.}} & & With sensor & Without sensor & & \textbf{Freq.} \\

  \hline

  59 & 4  & \multirow{7}{*}[7pt]{\boxplot[0.9]{59}{60}{62}{63}{65}} & \percent{0.08}
  & \percentright{0} & & 0 \\

  60 & 9  & &\percent{0.18}
  & \percentright{0} & & 0 \\

  61 & 8  & & \percent{0.16}
  & \percentright{0} & & 0 \\

  62 & 9  & & \percent{0.18}
  & \percentright{0} & & 0 \\

  63 & 11  & & \percent{0.22}
  & \percentright{0} & & 0 \\

  64 & 7 & & \percent{0.14}
  & \percentright{0} & & 0 \\

  65 & 2 & & \percent{0.04}
  & \percentright{0} & & 0 \\[2pt]
  \hline
  $\cdots$ & $\cdots$ &   &  $\cdots$ &  $\cdots$ &   &  $\cdots$ \\
  \hline

  81 & 0  &  & \percent{0}
  & \percentright{0.02} & \multirow{8}{*}[7pt]{\boxplot[1]{81}{84}{85}{86}{88}} & 1 \\
  82 & 0  &  & \percent{0}
  & \percentright{0.02} & & 1 \\
  83 & 0  &  & \percent{0}
  & \percentright{0.10} & & 5 \\
  84 & 0  &  & \percent{0}
  & \percentright{0.28} & & 14 \\
  85 & 0  &  & \percent{0}
  & \percentright{0.32} & & 16 \\
  86 & 0  &  & \percent{0}
  & \percentright{0.16} & & 8 \\
  87 & 0  &  & \percent{0}
  & \percentright{0.08} & & 4 \\
  88 & 0  &  & \percent{0}
  & \percentright{0.02} & & 1 \\
\hline

  & \multicolumn{3}{r}{ $ \textbf{p-value} < 2.2 \cdot 10^{-16}$} &\multicolumn{3}{l}{\footnotesize (Result of Welch's t-test)} \\
 &\multicolumn{3}{r}{mean (\scriptsize{Total evac. time}): \qquad \normalsize 61.86s} &\multicolumn{3}{l}{84.76s} \\
\hline

\end{tabular}
\caption[Analyzing the DensitySensor]{
  Comparison of total evacuation time frequencies of simulations with \cclass{DensitySensor} and without sensors.
  We conducted 50 simulations each.
  The simulations are all done with complete cognitive maps.
  }
\label{tab:sim_tjunction_complete_Density}
\end{table}

\section{Conclusion}
In this work we implemented a versatile and knowledge based routing framework for pedestrian dynamics.
With this framework we propose a adjustable method for emulating human knowledge, perception and decision making.
This work does not claim investigating and understanding the nature of human behavior, rather its goal is to create tools to ease the implementation of new behavioral models.

The framework consists out of three modules: the perception module (\cclass{Sensors} and \cclass{SensorManager}), the knowledge module (\cclass{Cogntivemap}) and the decision making module.
For the knowledge representation we proposed a simplified cognitive map which reduces the complexity of the model but represents all needed knowledge.
For the perception module we implemented a sensor structure and for the decision making we provide a local and a global optimization algorithm.
We showed that the sensors have a high impact on the simulation and are suitable for reproducing human behavior.
Especially the sensor module is highly extendable and could thereby fulfill even additional requirements later.
For all module the cognitive map is the central module to read information from or to write information into.
Due to the object oriented design and the modularization it is possible to exchange or adjust modules independently from each other.

There are several extensions and improvements which could be part of future works.
Some of them are related to problems which arose during the implementation or the analysis and some are suggestions for possible extensions.
One of the most important tasks would be a verification of simulations with empirical data.
This task is not just important for this routing framework but for the pedestrian dynamics simulations in general.
The next tasks are directly related to our new routing framework and suggestions for further improvements.

The perception module offers further extension possibilities.
We already implemented several sensors and showed their impact on the result of simulations.
For simulating certain situations and circumstances sensors are a good option.
That is why it is advisable to design and implement further sensors.
They enrich information and help to emulate realistic behavior.

The decision making should always use all knowledge which is available for the certain agent.
Additional the strategies could be mixed up to emulate more realistic behavior.

\end{document}